\documentstyle[12pt]{article}
\def\lromn#1{\uppercase\expandafter{\romannumeral#1}}

\begin{document}

\begin{flushright}
TU/96/514 \\
\end{flushright}

\vspace{12pt}

\begin{center}
\begin{Large}

\bf{
Quantum Dissipation in Open Harmonic Systems: 
Operator Solution
}
\end{Large}

\vspace{36pt}

\begin{large}
I. Joichi, Sh. Matsumoto, and M. Yoshimura

Department of Physics, Tohoku University\\
Sendai 980-77 Japan\\
\end{large}

\vspace{54pt}

{\bf ABSTRACT}
\end{center}

A finite number of harmonic oscillators coupled to infinitely many 
environment oscillators
is fundamental to the problem of understanding quantum dissipation 
of a small system
immersed in a large environment.
Exact operator solution as a function of time
is given to this problem, by using diagonalized dynamical variable
of the entire system, the small system plus the environment.
The decay law of prepared initial configuration is worked out
in greatest detail.
A clear separation of the exponential- and the power-law decay period
is made possible by our method.
Behavior of physical quantities at asymptotically late times can
be understood in terms of the overlap probability of the system variable
with the diagonal variable of the entire system.


\newpage

How a small system behaves under influence of a larger environment
has been investigated in a variety of approaches.
The simplest, yet the most fundamental model of quantum dissipation
is harmonic oscillator coupled to infinitely many oscillators that
make up a bath in a mixed state. Two powerful methods to analyze
this problem are the quantum Langevin equation \cite{q-dissipation in hos}, 
and the path integral approach \cite{q-dissipation in hos}.
In our previous work \cite{jmy-96-1} we used the path integral method
for the linear open system, and derived
the reduced density matrix of the system, by integrating out environment
variables.
In the present work we employ a direct operator approach.
Our new approach has an advantage, giving explicit operator solution
and thereby extends previous results \cite{jmy-96-1} 
to arbitrary initial states including correlation
between the system and the environment, and to non-thermal environments.

Consider a harmonic oscillator denoted by its coordinate variable $q$, 
interacting with bath oscillators
$Q(\omega )$'s, labeled by the continuous frequency $\omega $.
The Hamiltonian of the entire system
is assumed to be given by a sum of the kinetic term,
\( \:
\frac{1}{2}\, \left( \,p^{2} + \int d\omega P^{2}(\omega )\,\right)
\: \)
and the potential term,
\begin{equation}
V = \frac{1}{2}\, \left( \begin{array}{cc}
q & Q(\omega )
\end{array}
\right) \,{\cal V}\,
\left( \begin{array}{c}
q  \\ Q(\omega' )
\end{array}
\right) \,, \hspace{0.5cm} 
{\cal V} = \left( \begin{array}{cc}
\omega _{0}^{2} &  c(\omega' ) \\
c(\omega ) & \omega ^{2}\,\delta (\,\omega - \omega'\,)
\end{array}
\right) \,, 
\end{equation}
written in the matrix notation.
$c(\omega )$ describes the coupling of the variable $q$,
which we regard as the small system in question, 
to the environment variable $Q(\omega )$.
The potential part may be decomposed into the two,
\( \:
{\cal V} = {\cal V}_{0} + {\cal V}' \,, 
\: \)
where ${\cal V}_{0}$ consists of the environment part alone, 
\begin{eqnarray}
&& 
{\cal V}_{0} = 
\left( \begin{array}{cc}
0 & 0 \\
0 & \omega ^{2}\,\delta (\omega - \omega')
\end{array}
\right) \,, 
\\ &&
{\cal V}' = 
|{\cal S}_{1}\rangle \langle {\cal S}_{1}| - 
|{\cal S}_{2}\rangle \langle {\cal S}_{2}|
\,, \hspace{0.5cm}  
|{\cal S}_{i}\rangle \langle {\cal S}_{i}| = 
\langle {\cal S}_{i}|{\cal S}_{i} \rangle \,{\cal P}_{i} 
\,, \hspace{0.5cm} {\cal P}_{i}^{2} = {\cal P}_{i} \,, 
\\ &&
|{\cal S}_{1}\rangle = \left( \begin{array}{c}
\omega _{0}  \\
\frac{c(\omega )}{\omega _{0}}
\end{array}
\right) \,, \hspace{0.5cm} 
|{\cal S}_{2}\rangle = \left( \begin{array}{c}
0  \\
\frac{c(\omega )}{\omega _{0}}
\end{array}
\right) \,, 
\\ && 
\langle {\cal S}_{i}|{\cal S}_{i}\rangle  = 
\omega _{0}^{2}\,\delta _{i1} + \frac{1}{\omega _{0}^{2}}\,
\int_{\omega _{c}}^{\infty }\,d\omega \,2\omega \,r(\omega ) \,, 
\end{eqnarray}
where $\omega _{c}$ is the smallest value of the environment spectrum.
$r(\omega ) = c^{2}(\omega )/ (2\omega )$ 
characterizes the response of environment on the system. 
Our results in the present work hold for any $r(\omega )$ obeying
certain convergence property.

A typical question one frequently asks with regard to the system behavior
is how a pure state of the $q$ system
evolves in time in environment described by a mixed state such as
the thermal one of $Q(\omega )$'s.
We shall answer this by giving explicit time dependent operator
solution for $q(t) \,, p(t)$ that is written in terms of initial values,
\( \:
q_{i} \,, p_{i} \,, Q_{i}(\omega ) \,, P_{i}(\omega ) \,.
\: \)
With this one can clearly express correlators such as
\( \:
\langle q(t_{1})q(t_{2}) \rangle
\: \)
in terms of the distribution of these initial values in any state.
Closest to our present work is the classical work of Ullersma
\cite{ullersma}, with an important difference of the presence of a gap
in the environment spectrum, which yields different behaviors 
of the correlators.
The gapless case is applied  for instance to phonons in medium,
but in the case of our interest such as the unstable
particle decay in medium, a gap exists if the mass of the daughter particle
is finite.

The model itself, a harmonic system coupled to an infinite number 
of harmonic oscillators,
frequently appears in many idealized physical problems.
We utilize in the present investigation 
exact solution of the Hamiltonian eigenvalue problem to this system,
which might also be useful in other contexts.
Essentially the same solution of diagonalization is given 
in ref \cite{efimov-waldenfels} in a related model, 
but we use a modified form with an extra phase factor attached,
which makes our subsequent derivation transparant.

We first solve a related problem by seeking 
the eigenvector $|\Psi (\omega ) \rangle $
diagonalizing the matrix ${\cal V}$, 
using the unperturbed base $|\omega \rangle $,
\( \:
\left( \,{\cal V}_{0} - \omega ^{2}\,\right)\,|\omega \rangle = 0 \, ;
\: \)
\begin{eqnarray}
|\Psi (\omega )\rangle = \frac{1}{N(\omega )}\,|\omega \rangle 
- \frac{1}{{\cal V}_{0} - \omega ^{2} + i0^{+}}\,\left( \,
|{\cal S}_{1}\rangle \langle {\cal S}_{1}| - 
|{\cal S}_{2}\rangle \langle {\cal S}_{2}|
\,\right)\,
|\Psi (\omega )\rangle \,. \label{vector-ansatz} 
\end{eqnarray}
We have introduced the out-state boundary condition:
The boundary condition is determined by the prescription of the limit,
\( \:
i \lim_{\epsilon  \rightarrow 0^{+}}\,\epsilon \equiv i 0^{+} \,, 
\: \)
which extracts well-behaved solutions at $t \rightarrow + \infty $,
as in the Lippman-Schwinger equation.
This choice of the boundary condition is superficial and not dictated, 
to the extent that our problem is diagonalization by orthogonal
transformation of hermitian oscillator variables.
But we shall soon see how the solution with this form is useful to
solving our problem.

Our ansatz of the eigenvalue problem (\ref{vector-ansatz}) assumes that
there exists no isolated spectrum outside the continuos one.
An isolated spectrum, in the context of the problem raised here,
corresponds to a bound, stable state,
whose presence would imply that an initially prepared configuration 
of the system does not completely decay by interaction with environment. 
We exclude this case for our consideration.
Let us clarify under what condition the spectrum without any isolated
spectrum is realized.
The equation that governs the real part of the isolated
energy eigenvalue squared $\lambda $ is given by
\begin{equation}
f(\lambda ) = \lambda \,, \hspace{0.5cm} 
f(\lambda ) \equiv \omega _{0}^{2} + {\cal P}\,\int_{\omega _{c}}^{\infty }\,
d\omega \,\frac{2\omega\, r(\omega )}{\lambda - \omega ^{2}} \,.
\end{equation}
From the monotonic behavior of $f(\lambda )$ in the 
$\lambda < \omega _{c}^{2}$ region one readily derives
the condition of no zero in this region:
\begin{eqnarray}
f(\omega _{c}^{2 -}) > \omega _{c}^{2} \,, 
\label{condition of no real pole} 
\end{eqnarray}
with $^{-}$ indicating the limit from below.
Consistency further requires $\omega _{0} > \omega _{c}$.
When this condition (\ref{condition of no real pole}) is obeyed, 
the system energy $\omega _{0}$ moves,
with increase of the coupling $c(\omega )$, into the second Riemann
sheet of the complex $\omega $ plane having a cut at $\omega > \omega _{c}$.
The imaginary part of this location gives the decay rate of 
any initial configuration of the system, as will be shown later.

From the ansatz of $|\Psi (\omega )\rangle $ above we get equations for
\( \:
c_{i} = \langle {\cal S}_{i}|\Psi (\omega ) \rangle \,, \; (i = 1\,, 2)
\: \)
and thus eigenvectors,
\begin{eqnarray}
&& \hspace*{-1cm}
\left( \begin{array}{c}
c_{1}  \\ c_{2}
\end{array}
\right)
= \left( \begin{array}{cc}
- \,\frac{1}{\omega _{0}^{2}\,F(\omega - i 0^{+})} &  
1 - \frac{\omega _{0}^{2}}{\omega ^{2}} + \frac{1}{\omega _{0}^{2}
F(\omega  - i 0^{+})} 
\\
- \,1 + \frac{\omega _{0}^{2}}{\omega ^{2}} - \frac{1}{\omega _{0}^{2}
F(\omega  - i 0^{+} )} 
 & 2 - \frac{\omega _{0}^{2}}{\omega ^{2}} + \frac{1}{\omega _{0}^{2}\,
 F(\omega  - i 0^{+} )}
\end{array}
\right)^{-1}
\frac{c(\omega )}{\omega _{0}\,N(\omega )}
\left( \begin{array}{c}
1  \\ 1
\end{array}
\right) \,.
\end{eqnarray}
We introduced the analytic function $F(z)$ \cite{jmy-96-1};
\begin{eqnarray}
F(z) = \frac{1}{-\,z^{2} + \omega _{0}^{2} + 2\pi 
\overline{G}(z)}
\,, \hspace{0.5cm} 
\overline{G}(z) = 
\int_{-\infty }^{\infty }\,\frac{d\omega }{2\pi }\,
\frac{r(\omega )}{z - \omega }  \,,
\end{eqnarray}
with extended 
\( \:
r(- \omega ) = -\,r(\omega )
\: \)
for $\omega < 0$.
This function has cuts along the real axis, 
\( \:
\omega > \omega _{c} 
\: \)
and 
\( \:
\omega < -\,\omega _{c} \,. 
\: \)
We assume $\omega _{c} > 0$ such that there is a gap between the
two cuts.
The following discontinuity relation holds;
\begin{eqnarray}
F(\omega + i 0^{+}) - F(\omega - i 0^{+}) = i
2\pi \,r(\omega )F(\omega + i 0^{+})F(\omega - i 0^{+})
\equiv i 2\pi H(\omega ) \,, 
\end{eqnarray}
along this cut.

The previous function $f(\lambda )$ is related to this function by
\begin{eqnarray}
f(\lambda ) - \lambda = \Re F^{-1}(\sqrt{\lambda } + i0^{+}) \,.
\end{eqnarray}
An important property of $F(z)$ is that in the cut $z-$plane this function
is regular except on the real axis, and that it can be analytically
continued into the other Riemann sheets by the discontinuity formula.
In the second sheet $F(z)$ can be shown to have simple poles corresponding
to unstable states.
The location of these poles is given by
\begin{equation}
z ^{2} - \omega _{0}^{2} - 2\pi \,\overline{G}(z) 
+ 2\pi ir(z) = 0 \,. \label{spectrum zero} 
\end{equation}

Using the discontinuity formula, we may determine the the normalization,
leading to
\( \:
|N(\omega )|^{2} = 1 \,.
\: \)
Hence the overlap probability of the two vectors is 
\begin{eqnarray}
&&
|\,\langle 0|\Psi (\omega ) \rangle\,|^{2} = 2\omega \,H(\omega ) \,,
\\ &&
H(\omega ) =
\frac{r(\omega )}{(\,\omega ^{2} - \omega _{0}^{2}
- \Pi (\omega )\,)^{2} + (\pi r(\omega ))^{2}} \,, 
\end{eqnarray}
with 
\begin{equation}
\Pi (\omega ) = {\cal P}\,\int_{-\infty }^{\infty }\,d\omega '\,
\frac{r(\omega ')}{\omega - \omega '} \,.
\end{equation}
The eigenvector is then given by
\begin{eqnarray}
&&
\hspace*{-2cm}
|\Psi (\omega )\rangle = 
|\omega \rangle 
- c(\omega )F(\omega  - i 0^{+})\,|0\rangle 
+ c(\omega )F(\omega - i 0^{+})
\,\int_{\omega _{c}}^{\infty } d\omega' \frac{c(\omega' )}
{\omega'\,^{2} - \omega ^{2} + i 0^{+}}\,|\omega' \rangle 
\,. \label{out-state eq} 
\end{eqnarray}

It might seem odd due to an extra system degree of freedom, 
but it is possible to get inverted transformation to
express the original base in terms of the diagonal base thus derived;
\begin{eqnarray}
&&
|0\rangle = -\,\int_{\omega _{c}}^{\infty }\,d\omega \,
c(\omega )\,F^{*}(\omega - i 0^{+})\,|\Psi (\omega )\rangle \,, 
\\ &&
|\omega \rangle = 
|\Psi (\omega ) \rangle + c(\omega )\,
\int_{\omega _{c}}^{\infty }\,d\omega' \,c(\omega ')\,
\frac{F^{*}(\omega' - i 0^{+})}{\omega ^{2} - \omega'\,^{2} - i 0^{+}}
\,|\Psi (\omega' )\rangle \,.
\end{eqnarray}

How are these solutions for the state vectors related to our problem
of the operator diagonalization?
The crucial fact that can be verified straightforwardly is that
the complex coefficients for the out-state solution (\ref{out-state eq}) 
has a common phase of 
\( \:
e^{i\varphi (\omega )} = F(\omega - i 0^{+})/|\,F(\omega - i 0^{+})\,|
\: \).
This means that after taking out the phase factor 
$e^{i\varphi (\omega )}$ diagonalization via real orthogonal matix 
becomes possible, leading to hermitian diagonal variables 
$\overline{Q}(\omega )$. 
In the following computations we find it more convenient to use 
\( \:
\tilde{Q}(\omega ) \equiv e^{i\varphi (\omega )}\,\overline{Q}(\omega )
\: \)
by retaining these $\omega $ dependent phases.

With these remarks it should be evident that the canonical transformation
between the two descriptions exists;
\begin{eqnarray}
&& \hspace*{-1cm}
\tilde{Q}(\omega ) = Q(\omega ) - c(\omega )F(\omega - i 0^{+})\,
\left( \,q - 
\int_{\omega _{c}}^{\infty }\,d\omega '\,\frac{c(\omega ')}
{\omega '\,^{2} - \omega ^{2} + i 0^{+}}\,Q(\omega ')
\,\right) \,, 
\\ && \hspace*{-1cm}
\tilde{P}(\omega ) = P(\omega ) - c(\omega )F(\omega - i 0^{+})\,
\left( \,p - 
\int_{\omega _{c}}^{\infty }\,d\omega '\,\frac{c(\omega ')}
{\omega '\,^{2} - \omega ^{2} + i 0^{+}}\,P(\omega ')
\,\right) \,, 
\\ &&
q = -\,\int_{\omega _{c}}^{\infty }\,d\omega \,c(\omega )\,
F^{*}(\omega - i 0^{+})\,\tilde{Q}(\omega )  \,, 
\\ &&
p = -\,\int_{\omega _{c}}^{\infty }\,d\omega \,c(\omega )\,
F^{*}(\omega - i 0^{+})\,\tilde{P}(\omega ) \,, 
\\ &&
Q(\omega ) = \tilde{Q}(\omega ) + c(\omega )\,
\int_{\omega _{c}}^{\infty }\,d\omega' \,
\frac{c(\omega ')\,F^{*}(\omega' - i 0^{+})}
{\omega ^{2} - \omega '\,^{2} - i0^{+}}\,\tilde{Q}(\omega' ) \,, 
\\ &&
P(\omega ) = \tilde{P}(\omega ) + c(\omega )\,
\int_{\omega _{c}}^{\infty }\,d\omega' \,
\frac{c(\omega' )\,F^{*}(\omega' - i 0^{+})}
{\omega ^{2} - \omega '\,^{2} - i0^{+}}\,\tilde{P}(\omega' ) \,.
\end{eqnarray}
Assuming the canonical commutation for the original variables,
one can verify that diagonal variables obey the correct form of
the commutation relation;\\
\( \:
[ \,\overline{Q}(\omega ) \,, \overline{P}(\omega ')\,] =
i\,\delta (\,\omega - \omega '\,)
\: \)
etc.
It can be proved that with the specified phases of $\tilde{Q}(\omega ) \,, 
\tilde{P}(\omega )$ the original variables are all hermitian, as
required.

Operator diagonalization thus presented extends the work of 
\cite{efimov-waldenfels} only slightly: their model is simpler, with
only one dimensional projection operator involved 
in the interaction ${\cal V}'$, while ours has two dimensional projections.
It should however be clear that the diagonalization is possible
with any finite number of projections, with more complexities.

With this diagonal variable it is easy to write Heisenberg operator solutions
at any time $t$ in terms of the initial operator values; for instance
\begin{eqnarray}
&&
q(t) = -\,\int_{\omega _{c}}^{\infty }\,d\omega \,c(\omega )\,
F^{*}(\omega - i 0^{+})\,\tilde{Q}(\omega \,, t) \,, 
\\ &&
\tilde{Q}(\omega \,, t) = \cos (\omega t)\,\tilde{Q}_{i}(\omega ) +
\frac{\sin (\omega t)}{\omega }\,\tilde{P}_{i}(\omega ) \,.
\end{eqnarray}
The initial values $\tilde{Q}_{i} \,, \tilde{P}_{i}$ are then rewritten
in terms of the original variables.
After some straightforward calculation one finds that
\begin{eqnarray}
&&
\hspace*{-2cm}
q(t) = p_{i}\,g(t) + q_{i}\,\dot{g}(t) - \int_{\omega _{c}}^{\infty }\,
d\omega \,\sqrt{r(\omega )}\,\left( \,F^{*}(\omega - i 0^{+})\,e^{-\,i\omega t}
\,b_{i}(\omega ) + ({\rm h.c.})\,\right) \nonumber 
\\ &&
\hspace*{0.5cm} 
- \,\int_{\omega _{c}}^{\infty }\,d\omega ' \,
\int_{- \infty }^{\infty }\,d\omega \,H(\omega )\left( \,
\frac{\sqrt{r(\omega ')}}
{\omega ' - \omega + i0^{+}}\,e^{-\,i\omega t}\,b_{i}(\omega ')
+ ({\rm h.c.})\,\right) \,, \label{operator eq original}  
\end{eqnarray}
where 
\begin{eqnarray}
g(t) = 2\,\int_{\omega _{c}}^{\infty }\,d\omega \,H(\omega )\,\sin (\omega t)
\,, \label{g-def} 
\end{eqnarray}
and 
\( \:
b_{i}(\omega ) = (\,\sqrt{\omega }\,Q(\omega ) + i P(\omega )/\sqrt{\omega }\,)
/\sqrt{2}
\: \) is the annihilation operator for environment harmonic oscillators.

The formula for $q(t) $ and $p(t)$ is much simplified by introducing
\begin{eqnarray}
h(\omega \,, t) = \int_{0}^{t}\,d\tau \,g(\tau )\,e^{-\,i\omega \tau } \,, 
\end{eqnarray}
which is shown to be equal to
\begin{eqnarray}
h(\omega \,, t) = F(\omega - i 0^{+}) + \int_{-\infty }^{\infty }\,d\omega '
\,\frac{e^{-\,i (\omega - \omega ')t}}{\omega - \omega ' - i 0^{+}}\,
H(\omega ') \,. \label{h-identity} 
\end{eqnarray}
With this function one finds that
\begin{eqnarray}
q(t) = p_{i}\,g(t) + q_{i}\,\dot{g}(t) 
- \int_{\omega _{c}}^{\infty }\,
d\omega \,\sqrt{r(\omega )}\,\left( 
\,h^{*}(\omega \,, t)\,e^{-\,i\omega t}\,b_{i}(\omega ) + ({\rm h.c.})
\,\right) \,.
\label{q-correlator} 
\end{eqnarray}
Similarly with the aid of
\begin{eqnarray}
k(\omega \,, t) \equiv \int_{0}^{t}\,d\tau \,\dot{g}(\tau )e^{-\,i\omega 
\tau } = g(t)e^{- i\omega t} + i\omega h(\omega \,, t) \,, 
\end{eqnarray}
the momentum operator is given by
\begin{equation}
p(t) = 
p_{i}\,\dot{g}(t) + q_{i}\,\stackrel{..}{g}(t) 
- \int_{\omega _{c}}^{\infty }\,
d\omega \,\sqrt{r(\omega )}\,\left( 
\,k^{*}(\omega \,, t)\,e^{-\,i\omega t}\,b_{i}(\omega ) + ({\rm h.c.})
\,\right) \,. \label{p-correlator} 
\end{equation}

The basic formula (\ref{operator eq original}) was 
also derived by Ullersma \cite{ullersma}, 
who however did not exploit these solutions
in generality they really deserve. (The resonance approximation or
the local friction approximation is often made in deriving physical
consequences in that work.)
More importantly, the work of ref \cite{ullersma} assumed $\omega_{c} = 0$
(the gapless case). This special case does neither produce the power-law
decay nor the enhanced remnant as will be discussed subsequently.

It would be useful to clarify the physical significance of the function
$g(t)$ which is so basic in these formulas.
For this purpose let us derive a semi-classical equation of motion
by taking the ensemble average over the initial environment that has
no expectation value,
\( \:
\langle Q_{i}(\omega ) \rangle = \langle P_{i}(\omega ) \rangle = 0 \,.
\: \)
The expectation value of the system variable is then written in terms of
initial expectation values of the system variable:
\begin{equation}
\langle p(t)  \rangle = \langle p_{i} \rangle\,\dot{g}(t) + 
\langle q_{i} \rangle\,\stackrel{..}{g}(t) \,, 
\hspace{0.5cm} 
\langle q(t) \rangle = \langle p_{i} \rangle\,g(t) + 
\langle q_{i} \rangle\,\dot{g}(t) \,.
\end{equation}
Thus in the semi-classical picture the function $g(t)$ directly
controls time evolution.
Furthermore the initial values may be eliminated to give a closed equation 
of motion for the system variable:
\begin{eqnarray}
&&
\frac{d\,\langle p \rangle}{dt} = -\,\Omega (t)^{2}\,\langle q \rangle
- C(t)\,\langle p \rangle \,, \hspace{0.5cm} 
\frac{d\,\langle q \rangle}{dt} = \langle p \rangle \,,
\\ &&
\Omega ^{2}(t) = \frac{\dot{g}\stackrel{...}{g} - \stackrel{..}{g}^{2}}
{g\stackrel{..}{g} - \dot{g}^{2}} \,, \hspace{0.5cm} 
C(t) = \frac{\dot{g}\stackrel{..}{g} - g\stackrel{...}{g}}
{g\stackrel{..}{g} - \dot{g}^{2}} \,.
\label{time dependent friction} 
\end{eqnarray}
The quantities in this equation give the time dependent friction ($C(t)$) and
the time dependent frequency squared ($\Omega^{2}(t)$), 
incorporating environmental effects. 
The random force that appears in the quantum Langevin equation is
missing here, because we dropped the environment variable in taking
the ensemble average. 
Thus $g(t)$ describes an average behavior of the system variable
disregarding the random force from environment.

Either $g(t)$ or $\dot{g}(t)$ can be shown to follow 
integro-differential equation of the following form ($y = g \;$ or
$\dot{g}$),
\begin{eqnarray}
&&
\frac{d^{2}y}{dt^{2}} + \omega _{0}^{2}\,y + 2\,\int_{0}^{t}\,d\tau \,
\alpha _{I}(t - \tau  )\,y(\tau ) = 0 \,, 
\\ &&
\alpha _{I}(\tau ) = -\,\frac{i}{2}\,\int_{-\infty }^{\infty }\,d\omega \,
r(\omega )\,e^{-\,i\omega \tau } \,.
\end{eqnarray}
$\alpha _{I}(\tau )$ is the imaginary part of the kernel function that
appears in the influence functional \cite{feynman-vernon}.
Its significance on the system behavior
is that it incorporates integrated effects of the environment, the
non-local friction in general along with the frequency shift.
The two functions, $g(t)$ and $\dot{g}(t)$, 
differ in their boundary conditions:
\( \:
g(0) = 0 \,, \; \dot{g}(0) = 1 \,.
\: \)
The same differential equation also appeared in the path integral approach
\cite{qbm review}, \cite{jmy-96-1}.

The system variable has been determined in terms of the initial
operator values of both the system and the environment variables.
Dependence on the system initial values $p_{i} \,, q_{i}$ are given by
the function $g(t)$ as defined by (\ref{g-def}), which decreases first
exponentially and then finally by an inverse power of time, as can be
seen in the following way.
Using the discontinuity formula, one may rewrite the $\omega $ integration
containing 
\( \:
H(\omega ) = \left( \,F(\omega + i 0^{+}) - F(\omega - i 0^{+})\,\right)
/ (2\pi i)
\: \)
along the real axis into the $F(z)$ integration, the
complex $z$ running both slightly above and below the cuts.
The factor $\sin (\omega t)$ is replaced by $\Im e^{-\,i\omega t}$
in this procedure.
A half of this complex contour can be deformed into the second sheet,
and one thereby encounters simples poles in the second sheet, as
shown in Fig.1.
We assume for simplicity that there exists only a single pole 
in the nearby second sheet.
The intregral for $g(t)$ may then be expressed as 
the sum of the pole contribution (at $z = z_{0} $ with 
$ \Im z_{0} < 0$) in the second sheet and the contribution
parallel to the imaginary axis passing through $z = \omega _{c} $, both
in the first (\lromn 1) and in the second (\lromn 2)
sheet \cite{goldberger-watson}:
\begin{eqnarray}
&& \hspace*{-2cm}
g(\tau ) =
\Im \left( Ke^{-i\Re z_{0}\tau } \right)\,
e^{\Im z_{0}\tau } 
+
\Im \left[ \frac{e^{i\omega _{c}\tau }}{\pi }\int_{0}^{\infty }dy\,
e^{- y\tau }\left( \,F_{{\rm \lromn 1}}(\omega _{c} + iy) - F_{{\rm \lromn 2}}
(\omega _{c} + iy)\,\right) \right] \,, \label{g-integral} 
\end{eqnarray}
with
\( \:
K^{-1} = z_{0} - \pi \overline{G}'(z_{0} ) + i\pi r'(z_{0})\,.
\: \)

As seen from this formula, 
the pole contribution given by the first term describes
the exponential decay which usually lasts very long
during the most important phase of the decay period, while
the rest of contribution gives the power law decay at very late times;
$\propto t^{-\alpha -1}$.
The power $-\,\alpha -1$ is related to 
the threshold behavior of the response weight,
\( \:
r(\omega ) \:\propto  \: (\omega - \omega _{c})^{\alpha } \,.
\: \)
$\alpha  > - 1$ is required as a consistency of this approach;
convergence of the $\omega $ integration.
An example of $g(t)$, computed by the formula (\ref{g-def}), 
is shown in Fig.2, which clearly
exhibits both the exponential and the power law behavior.
The exponential period can be fitted precisely by the first pole term
of eq.(\ref{g-integral}) including the phase of oscillation, except
at very early times where quantum mechanical effect gives the non-decay
probability decreasing as 
\( \:
1 - O[t^{2}] \,.
\: \)
The exponent factor $-\,\Im z_{0} > 0$ gives the decay rate of any
initial configuration, and is related
to the constant friction coefficient $\eta $ in the usual treatment
of dissipation, namely, when quantum dissipation is assumed local
in time. This can be shown by computing the time dependent friction
$C(t)$ in eq.(\ref{time dependent friction}): when the non-pole term
is neglected in eq.(\ref{g-integral}),
\begin{equation}
C(t) = 2\gamma \,, \hspace{0.5cm} \Omega ^{2}(t) = \bar{\omega }^{2}
+ \gamma ^{2} \,,
\end{equation}
with an approximate form of
\( \:
g(t) = \frac{1}{\bar{\omega }}\,\sin (\bar{\omega }t)\,e^{-\,\gamma t}
\,.
\: \)
Both at very early and very late times these functions deviate
from the behavior in the pole approximation:
for instance $C(t) = O[t^{3}]$ as $t \rightarrow 0^{+}$.
At late times both of $\Omega ^{2}(t)$ and $C(t)$ decrease with
powers of time.

Thus at asymptotically late times the initial value dependence disappears
leading to
\begin{eqnarray}
&&
q(t) \:\rightarrow  \: -\,\int_{\omega _{c}}^{\infty }\,d\omega \,
\sqrt{r(\omega )}\,\left( \,F^{*}(\omega - i 0^{+})\,e^{-\,i\omega t}
\,b_{i}(\omega ) + ({\rm h.c.})\,\right) \,, 
\\ &&
p(t) \:\rightarrow  \: i\,\int_{\omega _{c}}^{\infty }\,d\omega \,\omega \,
\sqrt{r(\omega )}\,\left( \,F^{*}(\omega - i 0^{+})\,e^{-\,i\omega t}
\,b_{i}(\omega ) - ({\rm h.c.})\,\right) \,,
\end{eqnarray}
due to $h(\omega \,, \infty ) = F(\omega - i0^{+})$, which may be proved from
(\ref{h-identity}).
The system evolution is thus governed by the initial distribution of
the environment variables, $b_{i}(\omega ) \,, b_{i}^{\dag }(\omega )$,
with the following probability functions;
\begin{eqnarray}
r(\omega )\,|h(\omega \,, t)|^{2} \hspace{0.5cm} 
 {\rm for} \; q(t) \,, \hspace{0.5cm} 
r(\omega )\,|k(\omega \,, t)|^{2} \hspace{0.5cm}  {\rm for} \; p(t) \,.
\end{eqnarray}
that have limits of
\begin{eqnarray}
&&
r(\omega )|h(\omega \,, \infty )|^{2} = H(\omega ) = 
\frac{|\,\langle 0|\Psi (\omega ) \rangle\,|^{2}}{2\omega } \,, 
\\ &&
r(\omega )|k(\omega \,, \infty )|^{2} = \omega^{2}H(\omega ) = 
\frac{\omega }{2}\, |\,\langle 0|\Psi (\omega ) \rangle\,|^{2} \,.
\end{eqnarray}
Note that $H(\omega )$, hence the overlap probability
$|\,\langle 0|\Psi (\omega ) \rangle\,|^{2}$ has a universal character,
determined by general properties of the system and the environment alone,
irrespective of their particular initial states.
The overlap probability between the system $q$ variable and the diagonal
$\overline{Q}(\omega )$ respects the unitarity relation,
\begin{equation}
\int_{\omega _{c}}^{\infty }\,d\omega \,
|\,\langle 0|\Psi (\omega ) \rangle\,|^{2} = 
\int_{\omega _{c}}^{\infty }\,d\omega \,2\omega \,H(\omega ) = 1\,,
\end{equation}
which can be proved for any $r(\omega )$. This indicates that
the final state is described in terms of $\overline{Q} \,, \overline{P}$ 
variables alone.

The important function $H(\omega )$ takes almost the form of Breit-Wigner
function having a complex pole at $\omega = z$ of eq.(\ref{spectrum zero} ), 
especially in the weak coupling limit, $c(\omega ) \rightarrow 
0$. This leads to the approximation to neglect non-pole terms in 
(\ref{g-integral})
for practical applications. We however warn that some important
physics is missed in this approximation as will be made clear later.

Various physical quantities of the system under question can be
computed using our formulas written in terms of the initial values of
the system and the environment.
We consider the situation in which the environment is described by
some mixed state.
Let us assume for simplicity no correlation 
between the system and the environment at an initial time,
and furthermore that the initial environment density matrix is diagonal
with the Hamiltonian or the number operator of each $\omega $,
\( \:
n_{i}(\omega ) \equiv b_{i}^{\dag }(\omega )b_{i}(\omega ) \,,
\: \)
such that 
\( \:
\langle b_{i}(\omega )b_{i}(\omega ') \rangle = 0
\: \)
etc.
Various correlation functions are then evaluated;
for instance,
\begin{eqnarray}
&&
\langle q(t_{1})q(t_{2}) \rangle = -\,\frac{i}{2}\,g(t_{1} - t_{2})
+ \int_{0}^{t_{1}}\,d\tau \int_{0}^{t_{2}}\,ds\,g(t_{1} - \tau )
\alpha _{R}(\tau - s)g(t_{2} - s) \nonumber 
\\ && \hspace*{-1cm}
+\, g(t_{1})g(t_{2})\,\langle p_{i}^{2} \rangle + \dot{g}(t_{1})\dot{g}(t_{2})
\,\langle q_{i}^{2} \rangle + (\,g(t_{1})\dot{g}(t_{2})
+ \dot{g}(t_{1})g(t_{2})\,)\,\frac{1}{2}\, \langle q_{i}p_{i} +
p_{i}q_{i}\rangle \,, 
\\ &&
\alpha _{R}(\tau ) = \int_{\omega _{c}}^{\infty }\,d\omega \,
\langle 2n_{i}(\omega ) + 1 \rangle\,\frac{\cos (\omega t)}{2\omega }\,.
\end{eqnarray}
The first term of the correlator comes from the anti-symmetric part,
hence the commutator part $[\,q(t_{1}) \,, q(t_{2})\,]$, while the
rest from the symmetric part.
In this computation the following identity was used;
\begin{equation}
\hspace*{-1cm}
g(t_{1} - t_{2}) = g(t_{1})\dot{g}(t_{2}) - \dot{g}(t_{1})g(t_{2})
+ i\,\int_{\omega _{c}}^{\infty }\,d\omega\, r(\omega )
\left( \,h^{*}(\omega \,, t_{1})h(\omega \,, t_{2})\,e^{- i \omega 
(t_{1} - t_{2})} - ({\rm c.c.})\,\right) \,.
\end{equation}

Coincident time limits are evaluated from these, resulting in
\begin{eqnarray}
&& 
\langle q^{2}(t) \rangle = 
\int_{\omega _{c}}^{\infty }\,d\omega \,
\langle 2n_{i}(\omega ) + 1 \rangle\,r(\omega )\,|h(\omega \,, t)|^{2} 
\nonumber \\ &&
\hspace*{1cm} 
+ \,
g^{2}(t)\,\langle p_{i}^{2} \rangle + 
\dot{g}^{2}(t)\,\langle q_{i}^{2} \rangle + g(t)\dot{g}(t)\,
\langle p_{i}q_{i} + q_{i}p_{i} \rangle \,, 
\\ && 
\langle p^{2}(t) \rangle =
\int_{\omega _{c}}^{\infty }\,d\omega \,
\langle 2n_{i}(\omega ) + 1 \rangle\,r(\omega )\,|k(\omega \,, t)|^{2} 
\nonumber 
\\ &&
\hspace*{1cm} 
+\,
\dot{g}^{2}(t)\,\langle p_{i}^{2} \rangle 
+ \stackrel{..}{g}^{2}(t) \langle q_{i}^{2} \rangle + 
\dot{g}(t)\stackrel{..}{g}(t)
\langle p_{i}q_{i} + q_{i}p_{i} \rangle \,, 
\\ &&
\frac{1}{2}\, \langle q(t)p(t) + p(t)q(t) \rangle
= 
\int_{\omega _{c}}^{\infty }\,d\omega \,
\langle 2n_{i}(\omega ) + 1 \rangle\,r(\omega )\,h(\omega \,, t)
k^{*}(\omega \,, t) 
\nonumber \\ &&
\hspace*{0.5cm} +\,
\dot{g}(t)g(t)\,\langle p_{i}^{2} \rangle + \dot{g}(t)\stackrel{..}{g}(t)
\langle q_{i}^{2} \rangle + (\,\dot{g}^{2}(t) + g(t)\stackrel{..}{g}(t)\,)
\,\frac{1}{2}\, \langle p_{i}q_{i} + q_{i}p_{i} \rangle \,.
\end{eqnarray}
Asymptotic values are then
\begin{eqnarray}
&&
\langle q^{2}(\infty ) \rangle = 
\int_{\omega _{c}}^{\infty }\,d\omega \,
\frac{|\,\langle 0|\Psi (\omega ) \rangle\,|^{2}}{2\omega }\,
\langle 2n_{i}(\omega ) + 1 \rangle \,, 
\\ &&
\langle p^{2}(\infty ) \rangle = 
\int_{\omega _{c}}^{\infty }\,d\omega \,\frac{\omega }{2}\,
|\,\langle 0|\Psi (\omega ) \rangle\,|^{2}\,
\langle 2n_{i}(\omega ) + 1 \rangle \,.
\end{eqnarray}
More generally, for both $t_{1}$ and $t_{2}$ in the asymptotic late
time region,
\begin{eqnarray}
&&
\langle q(t_{1})q(t_{2}) \rangle \:\rightarrow  \:
\int_{\omega _{c}}^{\infty }\,d\omega 
\frac{|\,\langle 0|\Psi (\omega ) \rangle\,|^{2}}{2\omega }\,
\cos \omega (t_{1} - t_{2})\,\langle 2n_{i}(\omega ) + 1 \rangle \,, 
\\ &&
\langle p(t_{1})p(t_{2}) \rangle \:\rightarrow  \:
\int_{\omega _{c}}^{\infty }\,d\omega\,\frac{\omega }{2}\, 
|\,\langle 0|\Psi (\omega ) \rangle\,|^{2}\,
\cos \omega (t_{1} - t_{2})\,\langle 2n_{i}(\omega ) + 1 \rangle \,.
\end{eqnarray}
The other correlator vanishes:
\( \:
\langle q(t_{1})p(t_{2}) + p(t_{2})q(t_{1}) \rangle \rightarrow 0 \,.
\: \)

As an illustration and for comparison with the result in the path
integral, 
we take the initial environment in thermal state of temperature
$T = 1/\beta $ written by 
$Q(\omega )$ (and not diagonal $\tilde{Q}(\omega )$):
\begin{equation}
\langle 2n_{i}(\omega ) + 1 \rangle
= \coth (\frac{\beta \omega }{2}) \,.
\end{equation}
If we take for the initial system a gaussian state such as
a thermal one having a different temperature from the environment 
(which includes the ground state in the limit of $T \rightarrow 0$), 
our results here coincide with previous formulas \cite{jmy-96-1}.

In particular, we find it important to stress again that the asymptotic
time limit of the occupation number,
\begin{eqnarray}
n(\infty ) &\equiv& \frac{1}{2}\, \langle \frac{p^{2}(\infty )}
{\bar{\omega}} 
+ \bar{\omega}\,q^{2}(\infty ) \rangle - \frac{1}{2} 
\nonumber \\ 
&=& 
\frac{1}{2}\, \int_{\omega _{c}}^{\infty }\,d\omega \,\coth (\frac{\beta 
\omega }{2})\,(\bar{\omega} + \frac{\omega ^{2}}{\bar{\omega}})\,H(\omega )
- \frac{1}{2} \,, 
\end{eqnarray}
with $\bar{\omega }$ the real part of the pole position, 
contains the fraction not suppressed by the usual Boltzmann factor.
When the pole term dominates, or equivalently one approximates $H(\omega )$
by the Breit-Wigner function,
then the temperature dependent part of the occupation number defined by
\begin{equation}
n^{\beta  } = 
\int_{\omega _{c}}^{\infty }\,d\omega \,\frac{1}{e^{\beta \omega }
- 1}\,(\bar{\omega}  + \frac{\omega ^{2}}{\bar{\omega} })\,H(\omega ) \,, 
\end{equation}
has the factor 
\( \:
e^{-\,\bar{\omega }/T}
\: \)
at low temperatures.
But this approximation is not good at low temperatures.
Indeed, let us examine a typical example by taking the form of
\( \:
r(\omega ) = c\,(\omega - \omega _{c})^{\alpha } \,, 
\: \)
with $0 < \alpha < 1$ in the range of $\omega _{c} < \omega < \Omega $
($\Omega \gg \omega _{c}$) and with
\( \:
\bar{\omega } \gg  {\rm Max}\;(\,\omega_{c} \,, T\,)
\: \).
The result is 
\begin{equation}
n^{\beta  } \approx \frac{c}{\bar{\omega }^{3}}\,\Gamma (\alpha + 1)\,
e^{-\,\beta \omega _{c}}\,T^{\alpha + 1} \,.
\end{equation}
$\Gamma $ is the Euler's gamma function.

This shows that instead of the exponential suppression at low temperatures
what is left in medium after the decay has a power-law 
behavior of temperature dependence ($\propto T^{\alpha + 1}$).
An implication of this behavior to the unstable particle decay,
as discussed in our previous paper \cite{jmy-96-1},
is that the remnant fraction in thermal medium does not suffer from
the Boltzmann suppression factor at temperatures even much lower than
the mass of the unstable particle.
This is because the approximate Boltzmann-like equation is based 
on S-matrix elements computed on the mass shell, while the true
quantum mechanical equation may contain quantities off the mass shell.
Although implicit in these formulas (since $\omega_{c} \rightarrow 0$
limit is singular), the presence of a gap,
\( \:
\omega_{c} = 2 \times 
\: \)
(daughter mass) is critical to obtain the enhanced remnant fraction.

The corresponding behavior of $g(t)$ in this case is
\begin{eqnarray}
g(t) \approx -\,\frac{2c}{\bar{\omega }^{4}}\,\Gamma (\alpha + 1)\,
\frac{\cos (\,\omega _{c}\,t + \frac{\pi }{2}\,\alpha \,)}
{t^{\alpha + 1}} \,.
\end{eqnarray}
One can estimate the transient time $t_{*}$ from the exponential period to
the power period by equating the two formulas of $g(t)$ in their
respective ranges, to obtain
\begin{equation}
t_{*} \approx \frac{1}{\gamma }\,\ln \left( \frac{\bar{\omega }^{3}}
{2c\,\Gamma(\alpha + 1)\,\gamma ^{\alpha + 1}}\right)
 \,,
\end{equation}
with 
\( \:
\gamma = -\,\Im z_{0} \,
\: \)
the decay rate.

We finally mention briefly how the results derived by the operator solution
are related to the path integral results.
The gaussian system such as the one mentioned is characterized 
by a gaussian reduced density matrix
and is determined by the three quantities;
\( \:
\langle q^{2}(t) \rangle \,, \; \langle p^{2}(t) \rangle \,, \;
\frac{1}{2}\, \langle q(t)p(t)+p(t)q(t) \rangle \,.
\: \)
The results for these quantities given here agree with the results 
obtained by the path integral approach in our previous work \cite{jmy-96-1}.
This shows that the two approaches are equivalent whenever both
can give answers.
We shall give more details on this equivalence in our forthcoming paper.
The two approaches are complementary:
the path integral approach can cope with more general dynamical systems
than the harmonic one, while the operator approach, although limited to
the linear harmonic system, offers a simple interpretation in terms of
the overlap probability of the initial system variable, with extention
made possible to the general initial state.
Both should be useful.

\vspace{1cm}
\begin{center}
{\bf Acknowledgment}
\end{center}

This work has been supported in part by the Grand-in-Aid for Science
Research from the Ministry of Education, Science and Culture of Japan,
No. 08640341.


\newpage
\begin{center}
\begin{LARGE}
{\bf Figure Caption}
\end{LARGE}
\end{center}

\vspace{0.5cm} 
\hspace*{-1cm}
{\bf Figure 1}

Deformation of the complex contour for calculation of $g(t)$.

\vspace{1cm}
\hspace*{-1cm}
{\bf Figure 2}

Behavior of the function $g(t)$ for the Ohmic spectrum weight
\( \:
r(\omega ) = c\,(\,\omega - \omega _{c}\,)
\: \)
\\
for $\omega _{c} < \omega < \Omega $. Chosen parameters are
\( \:
\omega _{c} = 1 \,, \hspace{0.5cm} \Omega = 100 \,, \hspace{0.5cm} 
c = 0.01 \,, \hspace{0.5cm} \omega _{0} = 2.11 \,.
\: \)
Only the positive region of $g(t) > 0$ is shown for a technical reason, 
with clear signals of oscillation observed here.

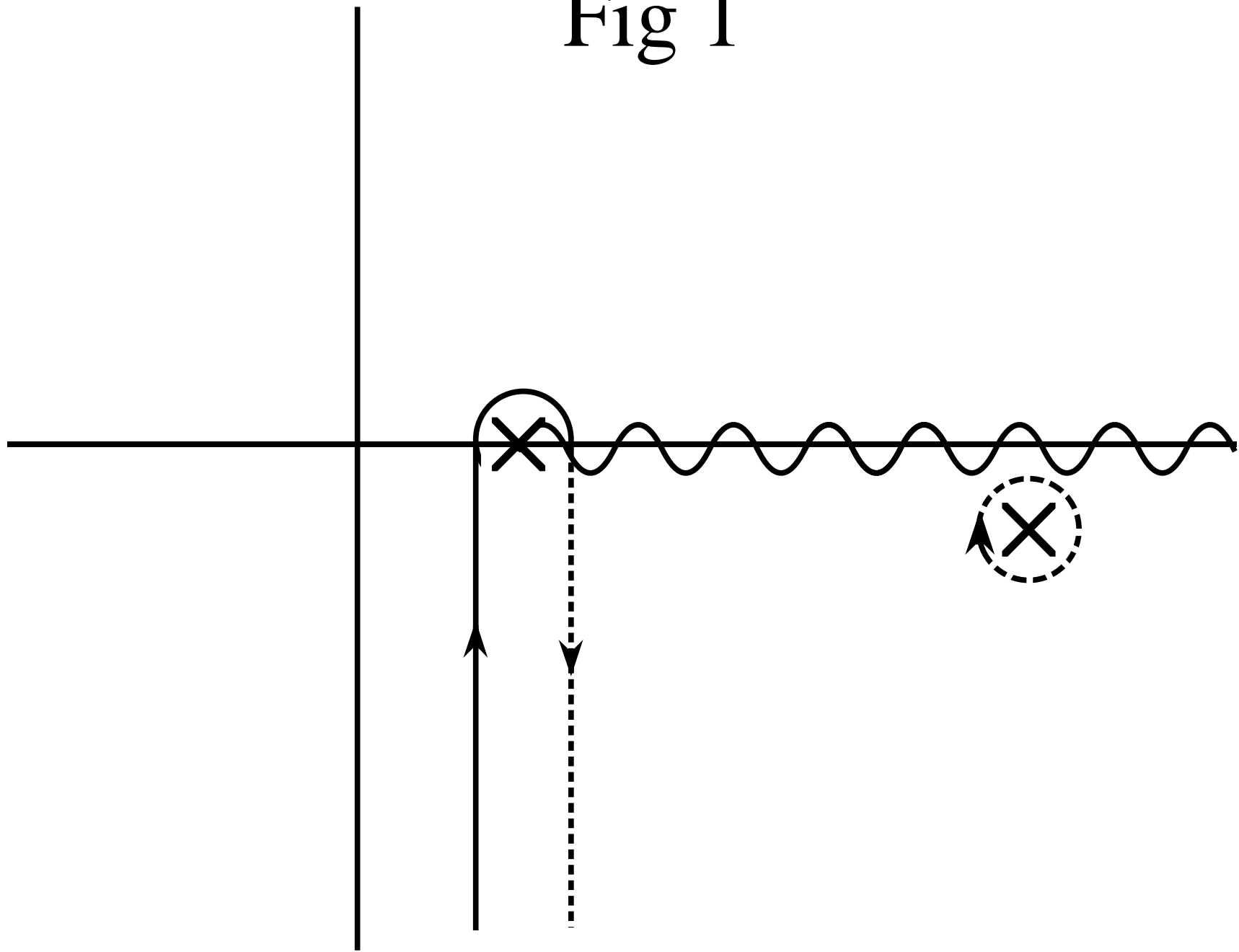
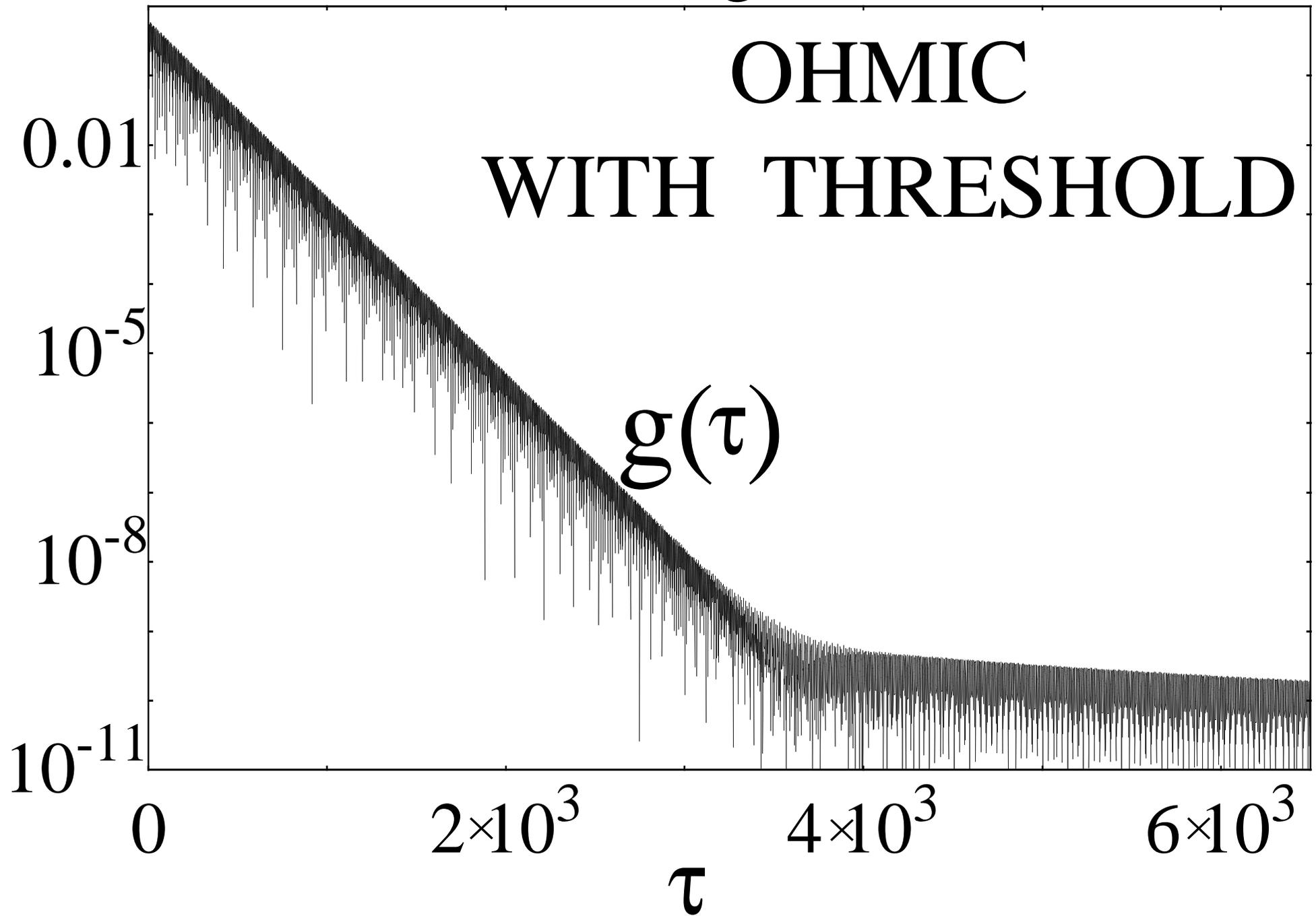
\end{document}